\documentstyle[twoside,epsfig]{article}

\catcode`\@=11
\long\def\@makefntext#1{
\protect\noindent \hbox to 3.2pt {\hskip-.9pt  
$^{{\eightrm\@thefnmark}}$\hfil}#1\hfill}		

\def\@makefnmark{\hbox to 0pt{$^{\@thefnmark}$\hss}}	
	
\def\ps@myheadings{\let\@mkboth\@gobbletwo
\def\@oddhead{\hbox{}
\rightmark\hfil\eightrm\thepage}   
\def\@oddfoot{}\def\@evenhead{\eightrm\thepage\hfil
\leftmark\hbox{}}\def\@evenfoot{}
\def\sectionmark##1{}\def\subsectionmark##1{}}

\oddsidemargin=\evensidemargin
\newcounter{sectionc}\newcounter{subsectionc}\newcounter{subsubsectionc}
\renewcommand{\section}[1] {\vspace{12pt}\addtocounter{sectionc}{1} 
\setcounter{subsectionc}{0}\setcounter{subsubsectionc}{0}\noindent 
	{\tenbf\thesectionc. #1}\par\vspace{5pt}}
\renewcommand{\subsection}[1] {\vspace{12pt}\addtocounter{subsectionc}{1} 
	\setcounter{subsubsectionc}{0}\noindent 
	{\bf\thesectionc.\thesubsectionc. {\kern1pt \bfit #1}}\par\vspace{5pt}}
\renewcommand{\subsubsection}[1] {\vspace{12pt}\addtocounter{subsubsectionc}{1}
	\noindent{\tenrm\thesectionc.\thesubsectionc.\thesubsubsectionc.
	{\kern1pt \tenit #1}}\par\vspace{5pt}}
\newcommand{\nonumsection}[1] {\vspace{12pt}\noindent{\tenbf #1}
	\par\vspace{5pt}}

\newcounter{appendixc}
\newcounter{subappendixc}[appendixc]
\newcounter{subsubappendixc}[subappendixc]
\renewcommand{\thesubappendixc}{\Alph{appendixc}.\arabic{subappendixc}}
\renewcommand{\thesubsubappendixc}
	{\Alph{appendixc}.\arabic{subappendixc}.\arabic{subsubappendixc}}

\renewcommand{\appendix}[1] {\vspace{12pt}
        \refstepcounter{appendixc}
        \setcounter{figure}{0}
        \setcounter{table}{0}
        \setcounter{lemma}{0}
        \setcounter{theorem}{0}
        \setcounter{corollary}{0}
        \setcounter{definition}{0}
        \setcounter{equation}{0}
        \renewcommand{\thefigure}{\Alph{appendixc}.\arabic{figure}}
        \renewcommand{\thetable}{\Alph{appendixc}.\arabic{table}}
        \renewcommand{\theappendixc}{\Alph{appendixc}}
        \renewcommand{\thelemma}{\Alph{appendixc}.\arabic{lemma}}
        \renewcommand{\thetheorem}{\Alph{appendixc}.\arabic{theorem}}
        \renewcommand{\thedefinition}{\Alph{appendixc}.\arabic{definition}}
        \renewcommand{\thecorollary}{\Alph{appendixc}.\arabic{corollary}}
        \renewcommand{\theequation}{\Alph{appendixc}.\arabic{equation}}
        \noindent{\tenbf Appendix \theappendixc #1}\par\vspace{5pt}}
\newcommand{\subappendix}[1] {\vspace{12pt}
        \refstepcounter{subappendixc}
        \noindent{\bf Appendix \thesubappendixc. {\kern1pt \bfit #1}}
	\par\vspace{5pt}}
\newcommand{\subsubappendix}[1] {\vspace{12pt}
        \refstepcounter{subsubappendixc}
        \noindent{\rm Appendix \thesubsubappendixc. {\kern1pt \tenit #1}}
	\par\vspace{5pt}}

\newcommand{\textlineskip}{\baselineskip=13pt}
\newcommand{\smalllineskip}{\baselineskip=10pt}

\def\eightcirc{
\begin{picture}(0,0)
\put(4.4,1.8){\circle{6.5}}
\end{picture}}
\def\eightcopyright{\eightcirc\kern2.7pt\hbox{\eightrm c}}

\def\abstracts#1#2#3{{
	\centering{\begin{minipage}{4.5in}\baselineskip=10pt\footnotesize
	\parindent=0pt #1\par 
	\parindent=15pt #2\par
	\parindent=15pt #3
	\end{minipage}}\par}} 

\newcommand{\bibit}{\nineit}

\renewenvironment{thebibliography}[1]
	{\frenchspacing
	 \ninerm\baselineskip=11pt
	 \begin{list}{\arabic{enumi}.}
	{\usecounter{enumi}\setlength{\parsep}{0pt}
	 \setlength{\leftmargin 12.7pt}{\rightmargin 0pt} 
	 \setlength{\itemsep}{0pt} \settowidth
	{\labelwidth}{#1.}\sloppy}}{\end{list}}

\newcounter{itemlistc}
\newcounter{romanlistc}
\newcounter{alphlistc}
\newcounter{arabiclistc}

\newcommand{\fcaption}[1]{
        \refstepcounter{figure}
        \setbox\@tempboxa = \hbox{\footnotesize Fig.~\thefigure. #1}
        \ifdim \wd\@tempboxa > 5in
           {\begin{center}
        \parbox{5in}{\footnotesize\smalllineskip Fig.~\thefigure. #1}
            \end{center}}
        \else
             {\begin{center}
             {\footnotesize Fig.~\thefigure. #1}
              \end{center}}
        \fi}

\newcommand{\tcaption}[1]{
        \refstepcounter{table}
        \setbox\@tempboxa = \hbox{\footnotesize Table~\thetable. #1}
        \ifdim \wd\@tempboxa > 5in
           {\begin{center}
        \parbox{5in}{\footnotesize\smalllineskip Table~\thetable. #1}
            \end{center}}
        \else
             {\begin{center}
             {\footnotesize Table~\thetable. #1}
              \end{center}}
        \fi}

\def\@citex[#1]#2{\if@filesw\immediate\write\@auxout
	{\string\citation{#2}}\fi
\def\@citea{}\@cite{\@for\@citeb:=#2\do
	{\@citea\def\@citea{,}\@ifundefined
	{b@\@citeb}{{\bf ?}\@warning
	{Citation `\@citeb' on page \thepage \space undefined}}
	{\csname b@\@citeb\endcsname}}}{#1}}

\newif\if@cghi
\def\cite{\@cghitrue\@ifnextchar [{\@tempswatrue
	\@citex}{\@tempswafalse\@citex[]}}
\def\citelow{\@cghifalse\@ifnextchar [{\@tempswatrue
	\@citex}{\@tempswafalse\@citex[]}}
\def\@cite#1#2{{$\null^{#1}$\if@tempswa\typeout
	{IJCGA warning: optional citation argument 
	ignored: `#2'} \fi}}

\def\pmb#1{\setbox0=\hbox{#1}
	\kern-.025em\copy0\kern-\wd0
	\kern.05em\copy0\kern-\wd0
	\kern-.025em\raise.0433em\box0}

\def\fnt#1#2{\footnotetext{\kern-.3em
	{$^{\mbox{\scriptsize #1}}$}{#2}}}

\def\fpage#1{\begingroup
\voffset=.3in
\thispagestyle{empty}\begin{table}[b]\centerline{\footnotesize #1}
	\end{table}\endgroup}

\def\runninghead#1#2{\pagestyle{myheadings}
\markboth{{\protect\footnotesize\it{\quad #1}}\hfill}
{\hfill{\protect\footnotesize\it{#2\quad}}}}
\font\tenrm=cmr10
\font\tenit=cmti10 
\font\tenbf=cmbx10
\font\bfit=cmbxti10 at 10pt
\font\ninerm=cmr9
\font\nineit=cmti9

\font\eightrm=cmr8

\def\qed{\hbox{${\vcenter{\vbox{			
   \hrule height 0.4pt\hbox{\vrule width 0.4pt height 6pt
   \kern5pt\vrule width 0.4pt}\hrule height 0.4pt}}}$}}

\def\w{{\rm w}}
\def\tr{{\rm tr}}
\def\k{{\bf k}}
\def\x{{\bf x}}
\def\p{{\bf p}}
\def\D{{\bf D}}
\def\E{{\bf E}}
\def\B{{\bf B}}
\def\v{{\bf v}}
\def\grad{\mbox{\boldmath$\nabla$}}
\def\bnum{\mbox{\bf\rlap{\#}{\kern1pt\#}}}
\def\Dslash{\mbox{\rlap{$D$}{\kern2pt/}}}

\begin{document}

\runninghead{Hot or Denser Matter}{Hot or Dense Matter}

\normalsize\textlineskip
\thispagestyle{empty}
\setcounter{page}{1}

~
\vspace{-1.0in}

\vspace*{0.88truein}

\fpage{1}
\centerline{\bf HOT OR DENSE MATTER\footnote{
   Talk given at DPF2000.
}}
\vspace*{0.37truein}
\centerline{\footnotesize PETER ARNOLD}
\vspace*{0.015truein}
\centerline{\footnotesize\it Department of Physics, University
of Virignia, P.O. Box 400714}
\baselineskip=10pt
\centerline{\footnotesize\it Charlottesville, Virginia 22904-4714,
USA}

\vspace*{0.21truein}
\abstracts{I review some of the properties of hot and/or dense
relativistic matter.}{}{}

\textlineskip			
\vspace*{12pt}			

\noindent
I've been asked to review hot and/or dense matter.  As a general rule, most
particle physicists aren't that familiar with the behavior of quantum field
theory at finite temperature and density.  We've all been introduced to how to
think about systems of a few relativistic particles with given energies, and
how to describe such systems with quantum field theory, but most of us
haven't had occasion to develop comparable experience and intuition about
systems of many particles at finite temperature or density.  I am therefore
going to forgo attempting a comprehensive review of recent results.
Instead, I will try to give a primer on relativistic hot or dense matter,
with a perhaps somewhat quirky choice of topics dictated by
my own areas of competence.

To begin, let me define some terms.  {\it Matter}\/ will mean gauge theories
(coupled to fermions and so forth).  As far as gauge theories go, I'll
restrict my attention to QCD and electroweak theory.  {\it Hot}\/ or
{\it dense}\/ will mean relativistically, and even ultra-relativistically,
hot or dense.

Ultra-relativistically hot QCD matter is matter at high density: if there
weren't any initially, they'd instantly appear due to pair creating collisions
of photons or gluons or whatever.  However, in this field, the adjective
``dense'' is used instead as a codeword to mean ``non-zero chemical potential
$\mu$.''  More specifically, dense typically refers to
non-zero chemical potential $\mu_{\rm B}$
for baryon number.  So dense QCD refers to the case where the numbers of quarks
and anti-quarks are significantly different.

\section{Hot Matter}

I'll begin by focusing just on hot matter, by which I mean $\mu=0$ or
close to it ($\mu \ll T$).  This is the situation of relevance to the
early Universe at temperatures above the QCD scale, when the asymmetry
between the number of quarks and anti-quarks was very small.

\subsection {Example: Electroweak baryogenesis}

One example of a situation where the physics of hot (but not ``dense'')
gauge theories is of interest is electroweak baryogenesis.\cite{bgenesis}
Electroweak baryogenesis is an attempt to explain the current dominance of
matter over anti-matter in our Universe in terms of electroweak physics.
It is based on the theoretical prediction that the standard model violates
baryon number at temperatures larger than the electroweak scale.  This is
a non-perturbative feature of the standard model, and it is caused by
certain types of non-perturbative fluctuations of electroweak gauge
fields.  Specifically, the change in baryon number $B$ turns out to be related
to the behavior of the electroweak gauge fields by what's known as an
anomaly equation:%
\footnote{
   For those of you familiar with chiral U(1) violation by
   the axial anomaly in QCD: this is similar.
}
\begin {equation}
   \Delta B = \# g_\w^2 \int d^4x \> \tr \, F_\w \tilde F_\w .
\end {equation}
I'm not actually going to need this equation for anything---just
remember that hot electroweak $B$ violation requires non-perturbatively large
fluctuations in the gauge fields.

\begin {figure}
\vbox{
   \begin {center}
      \epsfig{file=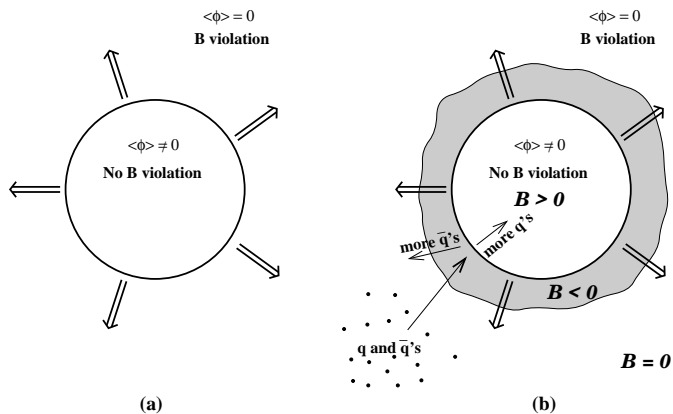,scale=.5}
   \end {center}
   \caption
       {%
        A cartoon of electroweak baryogenesis.
       \label{fig:bgenesis}
       }%
}
\end {figure}

It turns out that the probability of such non-perturbative fluctuations is
exponentially sensitive to the mass of the W and so is exponentially
sensitive to the expectation value $\langle\phi\rangle$ of whatever plays
the role of the Higgs, since that's where the W's mass comes from.
At the present time, with $\langle\phi\rangle$ of order the weak scale,
the rate for electroweak baryon number violation is predicted to be so
tiny that you wouldn't see it even if your proton decay experiment was
the size of the observable universe and ran for 15 billion years.
However, in the very early universe, $\langle\phi\rangle$ was effectively
zero, and $B$ violation is predicted to have been large.  Depending on the
details of the Higgs sector, there could have been an abrupt, supercooled,
first-order phase transition between the original, hot, symmetry-restored
($\langle\phi\rangle=0$) phase of electroweak theory and the current
symmetry-broken phase ($\langle\phi\rangle\not=0$).  This transition is
the focal point of scenarios of electroweak baryogenesis, a cartoon for
which is shown in Fig.\ \ref{fig:bgenesis}.
A supercooled first-order transition would proceed by the nucleation and
expansion of bubbles of the symmetry-broken phase.  Such a bubble is shown
in Fig.\ \ref{fig:bgenesis}a.  As the bubble expands, it will run into
quarks and anti-quarks present in the hot plasma that filled the Universe
at that time.  Because of CP violating interactions with the bubble wall,
it is possible for quarks to be able to penetrate the bubble wall slightly
more often than anti-quarks.  This results in an anti-quark excess building
up in a shell outside the wall (and a quark excess inside) as depicted in
Fig.\ \ref{fig:bgenesis}b.
Baryon number violation processes in that shell
attempt to locally restore chemical equilibrium by converting those
excess anti-quarks into quarks, and so increasing the baryon number of
the Universe.  There is no balancing conversion of
the excess quarks {\it inside}\/ the bubble wall because $B$ violation
is not significant in the symmetry-broken phase, as I asserted earlier.
So, in net, a baryon number $B>0$ for the universe is generated.

This scenario generates all sorts of interesting theoretical problems to work
on.  For example, one would like to know how to compute the
rate of non-perturbative processes, such as $B$ violation, in hot gauge
theory.  As another example, many aspects of the bubble wall growth
depend on hydrodynamics of the quark-gluon-W-Z-{\it etc.}\ plasma.
The parameters of hydrodynamics are known as transport
coefficients---viscosities, diffusion constants, {\it etc.}---and one would
like to know how to compute them from first principles.

The rate of $B$ violation is a non-perturbative problem, and the
usual way to attack non-perturbative problems from first principles is
to make lattice simulations.  However, there is a difficulty
here: standard lattice methods for quantum field theory are based
on simulations in imaginary time, and the quantities of interest here
live in real time!  This is potentially an extremely serious difficulty,
and I'll discuss later how it is overcome.

\subsection {Hot matter at small coupling}

The example of electroweak baryogenesis shows that there are interesting
questions to be addressed even in {\it weakly}-coupled hot field
theories, such as hot electroweak theory.  And, in studying relativistically
hot plasmas more generally, it seems a good idea to try to understand
something you might at first think was relatively simple---the weak
coupling limit.  So, for the rest of my talk, I'll specialize to
very hot QCD and Electroweak theory {\it at small coupling}, by which
I mean that $g(T) \ll 1$, where $g(T)$ is
the running coupling constant
at the
scale of the temperature $T$.  Also, for simplicity (and because it's
plenty complicated enough), I will focus on hot physics in systems that
are at or near equilibrium.

\subsubsection {Isn't small coupling trivial?}

\begin {figure}
\vbox{
   \begin {center}
      \epsfig{file=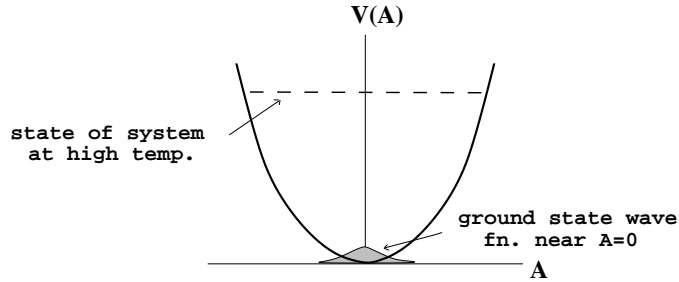,scale=.5}
   \end {center}
   \caption
       {%
       A slightly anharmonic oscillator.
       \label{fig:highT}
       }%
}
\end {figure}

You might think the small coupling limit is trivial, though possibly tedious,
to analyze: you just lock a theorist up in a closet with some paper and have
them do perturbation theory by computing a bunch of Feynman diagrams to
whatever order is desired.  At high temperature, however, small coupling
does {\it not} mean perturbative.  A simple way to understand this is to
think not of field theory but simple quantum mechanics.  Consider a
slightly anharmonic oscillator, depicted in fig.\ \ref{fig:highT},
with potential energy of the form
\begin {equation}
   V(x) \sim \omega_0^2 x^2 + g^2 x^4 ,
\end {equation}
and suppose that $g^2$ is very, very small.
In the ground state, the wave function will be localized around $x=0$,
and one will be able to treat $g^2 x^4$ as a perturbation compared to
$\omega_0^2 x^2$.  But now consider the state of the system at high
temperature.  The larger the temperature, the larger the energy, in
which case the larger the values of $x$ the system will be able to
probe.  For large enough $x$, a quartic $g^2 x^4$ will always dominate
over a quadratic $\omega_0^2 x^2$, and so $g^2 x^4$ cannot be treated
as a perturbation to $\omega_0^2 x^2$.

The physics of the slightly anharmonic oscillator becomes
non-perturbative at large $T$, but it also becomes {\it classical}.
The state of the system corresponds to very high level numbers,
and so is classical by the correspondence principle.
This is important: As I'll discuss later, it is the classical
nature of the non-perturbative physics that makes it possible to do
real-time lattice simulations for hot gauge theories.

Now note that if we fixed $T$ but varied $\omega_0$, and took
$\omega_0 \to 0$, we'd again get to a non-perturbative situation where
$g^2 x^4$ cannot be treated perturbatively compared to $\omega_0^2 x^2$.
This observation allows us to see how this quantum mechanical example
translated to field theory.  Hot gauge theory can be thought of as
coupled QM oscillators corresponding to each Fourier mode $\k$,
with corresponding natural frequencies $\omega_k \sim k$.
In QM, we see that $\omega_0 \to 0$ at fixed $T$ leads to non-perturbative
physics.  In hot gauge theory, $\omega_k \sim k \to 0$ will
analogously lead to
non-perturbative physics.  It is therefore the low-momentum (long wavelength)
modes of hot gauge theory that have non-perturbative fluctuations at
high temperature.  It turns out that the momentum scale at which the
physics becomes non-perturbative is $k \sim g^2 T$.

\subsubsection {Hey, what about deconfinement?}

I've asserted that there is non-perturbative physics in very hot gauge
theories.  You might wonder how this is consistent with the picture that
QCD deconfines at high temperature and, because of
asymptotic freedom, can be treated as a weakly-interacting gas of quarks
and gluons.  The origin of deconfinement is something that you learned
from Jackson in graduate school: static electric fields are Debye
screened in plasmas.  For an Abelian theory, the Coulomb potential $g^2/r$
between two static test charges is modified by Debye screening to
a Yukawa potential $g^2 e^{-m_{\rm D} T}/r$.  In non-Abelian theories,
a similar exponential screening of the potential occurs, screening away
the linear long-distance potential that is responsible for confinement.
In the ultra-relativistic limit, the inverse screening length
$m_{\rm D}$ turns out to be of order $g T$.

Static {\it magnetic}\/ fields, however, are not screened in plasmas.
This is why it's possible, for instance, for the Sun and the galaxy
to have magnetic fields.  It is these unscreened magnetic fields which
can produce non-perturbative physics.

It's worth mentioning, for the sake of later discussion, that plasmas
{\it do}\/ resist {\it dynamical}\/ magnetic fields.  This is due to
what's known in introductory physics as Lenz's Law: conductors resist
changes in magnetic field.  Plasmas are conductors, and changes in the
magnetic field create electric fields, which then create currents in the
plasma, which create magnetic fields opposing the original change.

I have now introduced you to a hierarchy of momentum scales in weakly coupled
ultra-relativistic plasmas:
\medskip
\begin {center}
\begin {tabular}{cl}
  $T$ & energy/momentum of typical particles in the plasma; \\
  $gT$ & inverse Debye screening length; \\
  $g^2 T$ & non-perturbative magnetic fluctuations. \\
\end {tabular}
\end {center}
\medskip
Other scales of interest turn out to
include the mean free path for color-randomizing
collisions, which is order $g^2 T \ln(1/g)$, and the mean free path for
large-angle scattering, which is $g^4 T \ln(1/g)$.
The trick to studying hot physics turns out to be to
find and exploit appropriate
effective theories which describe the physics at each scale of interest,
and systematically match the parameters of those effective theories
to the original gauge theory.

\subsubsection {Small coupling expansion is NOT the loop expansion}

Even though perturbation theory fails at high temperature, there is no
reason that one can't Taylor expand the result for physical quantities
in powers of the small coupling $g$.  For instance, the free energy of
a hot non-Abelian gauge theory has an expansion of the form\cite{braaten}
\begin {equation}
   F = T^4\left[\# + \# g^2 + \# g^3 + \# g^4
       + \#g^5\left(\ln{1\over g} + \#\right)
       + \#g^6\left(\ln{1\over g} + \bnum\right) + \cdots \right] .
\label {eq:F}
\end {equation}
Here, the $g^0$ term is the ideal gas result, and the other terms are
corrections due to interactions.  The fact that this is not simply
the loop expansion is manifest from the appearance of odd powers of $g$,
as well as logs of $1/g$.  The terms through $g^5$ can in fact be
calculated by pencil and paper using appropriately resummations of perturbation
theory, but the $g^6$ term denoted by the boldface $\bnum$ is the
leading contribution to the free energy from truly non-perturbative
magnetic physics.  It can't be calculated with pencil and paper.

We are used to seeing logs of ratios of physical scales in zero
temperature results.  For instance, the transverse distribution for
producing W's in a collider has logs of $p_{\rm T}/M_{\rm w}$.
In the previous section, I discussed how relevant physical scales at
finite temperature are $T$, $gT$, and $g^2T$.  It is logs of the ratios
of such scales that produce the logs of $1/g$ in (\ref{eq:F}).
[Ratios of these scales are also responsible for the odd powers of
$g$ in (\ref{eq:F}).]

The free energy is an example where non-perturbative physics doesn't
show up until high order in the small-coupling expansion (which is why
it remains useful to think of the plasma as a weakly-interacting gas of
quarks and gluons).  There are other quantities, such as baryon number
violation, that are intrinsically non-perturbative.  The baryon
number violation rate per unit volume turns out to have a
small-coupling expansion of the form\cite{Bform}
\begin {equation}
   \Gamma_{\rm B} = \bnum g_w^{10} T^4 \left(\ln{1\over g} + \#\right) + \cdots ,
\end {equation}
where $\bnum$ is a coefficient that cannot be computer with pencil and
paper.

\begin {figure}
\vbox{
   \begin {center}
      \epsfig{file=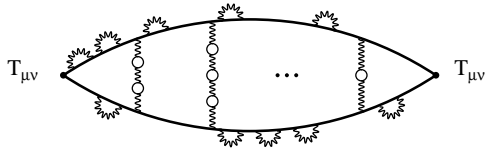,scale=.3}
   \end {center}
   \caption
       {%
       Example of the infinite class of diagrams that contribute to
       the leading-order result for inverse shear viscosity.
       \label{fig:shear}
       }%
}
\end {figure}

To emphasize the difference between the small coupling expansion and
the loop expansion, here's the small coupling expansion for
the inverse of the shear viscosity, which is\cite{shear}
\begin {equation}
   \eta^{-1} = \# g^4 T^{-3} \ln{1\over g} + \cdots .
\end {equation}
Diagrammatically, the leading term comes from an {\it infinite}\/ subset
of perturbative diagrams, an example of which is shown in
fig.\ \ref{fig:shear}.

\subsection{Effective Theories}

  Summing up infinite classes of diagrams by hand, such as those suggested by
fig. \ref{fig:shear}, is a pain in the butt, and is also of little use
if your ultimate interest is calculating something non-perturbative.
The most efficient way to proceed is to instead figure out {\it effective}\/
theories of the long-distance physics, which can be matched to the
short-distance physics of the original theory.

If one is interested in non-dynamical questions about hot, equilibrium
gauge theories---that is, if one doesn't care about time dependence---then
analyzing the system in imaginary time is just as good as analyzing it in
real time.  The use of effective theories appropriate to imaginary-time
analysis has a long history and is well understood and well under control.
I won't dwell on it here, except in a footnote.%
\footnote{
   At momentum scales well below $T$, the effective theory is
   3 dimensional (as opposed to 3+1 dimensional) gauge theory together
   with an adjoint scalar field $A_0$, and there are no fermions in the
   effective theory.  At momentum scales well below $gT$, the $A_0$
   decouples due to Debye screening, and the effective theory is just
   pure 3 dimensional gauge theory (no fermions).
}

Over the past few years, progress has been made in understanding the sequence
of effective theories that efficiently describe dynamical questions (baryon
number violation, transport coefficients, ...) at different scales.  As
suggested earlier, long distance physics is classical at high temperature.
The appropriate effective theory for distance scales large compared to $1/T$
({\it i.e.}\ momenta
small compared to $T$) can be described in classical terms by
what is known as a Boltzmann-Vlasov equation.  The state of the system at any
time is represented by two types of classical functions.  The first,
$n_i(\x,\p,t)$, represents the density in phase space of particles of a given
type $i$ (specifying flavor, color, and spin) at position $\x$ carrying
momentum $\p$.  The second, $A(\x,t)$, represents the configuration of the
${\it long{-}wavelength}$ modes of the gauge fields.
(Gauge field quanta with
short wave-lengths are treated as particles and incorporated into the
$n_i$'s.)  The equation that describes the time evolution is simply
a Boltzmann equation, which describes the effects on the densities
$n_i$ due to (1) free streaming in the background of the long-wavelength
field $A$, and (2) changes in $\p$ due to collision among each other.
Schematically, the equation looks like
\begin {equation}
   (D_t + \v\cdot\D) n + g(\E+\v\times\B)\cdot\grad_\p n
   = \mbox{collision term} .
\end {equation}
The first term is a convective derivative (which tells you that
$n$ can change at a given $\x$ simply because the particles' momenta
carried them away to a different $\x$)  The second term accounts for
the force on the particles due to the long wavelength gauge fields.
The collision term, if one were to write it out, involves expressions like
\begin {equation}
   \int_{234} |{\cal M}_{12\to34}|^2 n_1 n_2 (1\pm n_3)(1\pm n_4) ,
\end {equation}
where ${\cal M}_{12\to34}$ is, for example, a $2\to2$ scattering amplitude,
and the $(1\pm n)$'s are final-state Bose enhancement or Fermi blocking
factors.  Finally, one also needs the equation for the time evolution of
the long-wavelength gauge fields, which is just Maxwell's equation,
\begin {equation}
   D_\mu F^{\mu\nu} = j^\nu = \int_\p g v^\mu \, n(\p) ,
\end {equation}
with sources provided by the particles described by the $n$'s.
These two, simple, physically motivated  equations reproduce what are
known as ``hard thermal loops,'' which was historically the original analysis
of the effective dynamics at long distances, based on a difficult analysis
of resumming Feynman diagrams.

At yet larger distance scales, large compared to the scale
$g^4 T \ln(g^{-1})$ which turns out to characterize the mean-free
path for large-angle scattering, the appropriate effective theory
becomes linearized hydrodynamics.  (The Boltzmann-Vlasov effective
theory is still valid, but, like the original quantum field theory,
is no longer an efficient description for calculations.)

There is an important unresolved issue in the use of effective theories
for studying dynamical properties.  Unlike the non-dynamical case, it is
somewhat unclear how to improve the validity of these effective theories
beyond leading order in powers of $g$.  (At least, nobody has ever done
a systematic calculation of a physical quantity beyond leading order.)

\subsection {Example: numerical simulations of the B violation rate}

The nice thing about a classical theory is that, in principle, it's
straightforward to simulate it in real time.%
\footnote{
  In practice, there are all sorts of things that can hang you up,
  but I won't discuss them here.
}
\ You discretize space
into a lattice, pick initial configurations from a thermal ensemble,
evolve each configuration forward in time using the classical
equations of motion, and then measure whatever you wanted to measure.
The classical equations discussed previously (as well as some even simpler
effective theories I don't have time to discuss) have been used to simulate
the rate of baryon number violation.  Doing this a variety of different
ways with a variety of variations on the classical effective theory,
various subsets of Moore, B\"odeker, Rummukainen, Hu, and M\"uller have
found\cite{moore}
\begin {equation}
   \Gamma_{\rm B} \simeq
   \pmatrix{25\pm2\cr 25\pm7\cr 29\pm6} \alpha_\w^5 T^4
\end {equation}
for the Standard Model with $\alpha_\w \simeq 1/30$.  [I have not
explicitly shown the $\ln(1/\alpha_\w)$ factors but have incorporated
them into the numbers.]

\section {Dense Matter}

I'll now switch to a discussion of dense matter
(sometimes also hot, sometimes not) --- that is, matter at
non-zero chemical potential for baryon number.
I should disclose that I am not on expert at this subject, and there
were many people at the DPF conference much more knowledgeable than I.

Even {\it non}-dynamical questions about dense matter are
problematical for lattice simulations.
The problem is that the imaginary-time path integral has the form
\begin {eqnarray}
&&
   \int [{\cal D}A]\>[{\cal D}\psi] [{\cal D}\bar\psi]\>
   \exp[-S_{\rm YM} - \bar\psi(\Dslash + m + i\mu\gamma_0)\psi]
\nonumber\\ && \hspace{8em}
   =
   \int [{\cal D}A]\>
   \exp[-S_{\rm YM}] {\det}^N(\Dslash + m + i\mu\gamma_0) ,
\end {eqnarray}
and, because of the $i\mu\gamma_0$, the integrand on the right-hand
side is not positive real.  That means that the integrand can't be
interpreted as a probability measure on which to base a Monte Carlo
algorithm.

It's interesting to note that, despite the $i\mu\gamma_0$, the determinant
does turn out to be positive real for (a) two-color QCD,\cite{kogut}
or (b) non-zero
isospin chemical potential instead of non-zero baryon number chemical
potential.\cite{stephanov}
These unrealistic
cases should provide nice opportunities for numerical
simulation to test theoretical attempts to analyze gauge theories at
non-zero chemical potential.

\subsection{Example: The phase diagram of hot, dense QCD}

\begin {figure}
\vbox{
   \begin {center}
      \epsfig{file=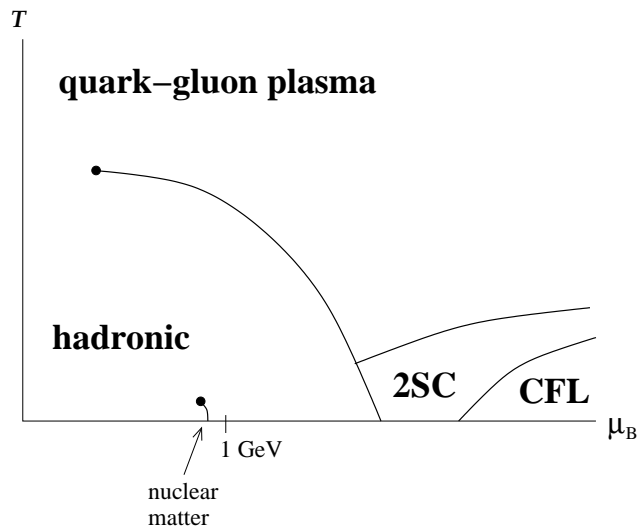,scale=.5}
   \end {center}
   \caption
       {%
       Educated guess at phase diagram of hot and dense
       QCD.\protect\cite{krishna}
       \label{fig:QCDphases}
       }%
}
\end {figure}

One educated guess as to what the phase diagram of hot, dense QCD might
look like is shown in fig.\ \ref{fig:QCDphases}, which I stole from
a conference talk of Krishna Rajagopal's.\cite{krishna}
The phases marked
2SC and CFL are color superconducting phases in which a color-breaking
diquark condensate $\langle qq\rangle \not=0$ forms.
2SC denotes a
2-flavor (u,d) color superconductor in which the SU(3) color gauge
group is only partly broken and in which chiral symmetry is unbroken.
CFL (an acronym for Color Flavor Locking) denotes a 3-flavor
(u,d,s) color superconductor in which chiral symmetry {\it is} broken.

Readers may well be interested to understand what to make of the claims
in the press that CERN has already discovered the quark-gluon plasma
region of this diagram.  Readers might be interested to know what the
prospects are for RHIC and the LHC
to explore, verify, or falsify any of the features of
fig.\ \ref{fig:QCDphases}.  These are interesting and important questions.
They are also complicated and intricate questions.
Now recall that earlier in this talk, I said I would focus on gauge
theories at weak coupling.  So, for the sake of consistency,
I sadly but clearly have no choice but to ignore
fig.\ \ref{fig:QCDphases} altogether and to move on.

\medskip

Well, perhaps that's a bit hasty...

\subsection{Dense QCD at weak coupling}

\begin {figure}
\vbox{
   \begin {center}
      \epsfig{file=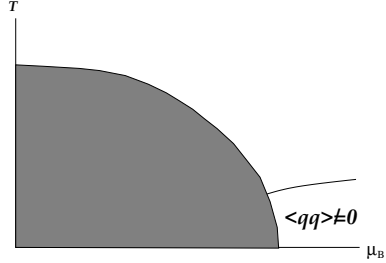,scale=.3}
   \end {center}
   \caption
       {%
         The phase diagram in the asymptotic, small-coupling region (unshaded).
         I have not differentiated between the different types of
         superconducting phases, marked here as $\langle qq\rangle \not= 0$.
       \label{fig:QCDphases2}
       }%
}
\end {figure}

As depicted in fig.\ \ref{fig:QCDphases2}, there is a portion of
the phase diagram which is at asymptotically
large values of $\mu$ [or $T$], and so the running coupling $g(\mu)$
[or $g(T)$] evaluated at the relevant physical scale is small, due to
asymptotic freedom.  Here, one should in principle be able to do a
semi-rigorous analysis of the problem from first principles.
That makes a good, solid check of educated guesses about the more
physically relevant parts of the diagram, now hidden by the shading
in fig.\ \ref{fig:QCDphases2}.

The basic principle behind the BCS theory of superconductivity is that,
at finite density and zero temperature, attractive interactions between
charged fermions lead to superconductivity, even if those attractive
interactions are arbitrarily weak.  The attraction causes the charged
fermions to bind into charged Cooper pairs, and those Cooper pairs
condense into a Bose condensate.  In particle physics language, the Cooper
pairs play the role of Higgs particles, the Cooper pair field $\psi\psi$ then
acquires a VEV, and that VEV gives mass to the photon.  A mass means a Yukawa
potential, which is why magnetic fields fall of exponentially quickly
inside a superconductor.  This is the Meisner effect.
In our case, the relevant fermions are quarks, the charge of interest is
the color charge, and the corresponding ``photons'' are gluons.

\begin {figure}
\vbox{
   \begin {center}
      \epsfig{file=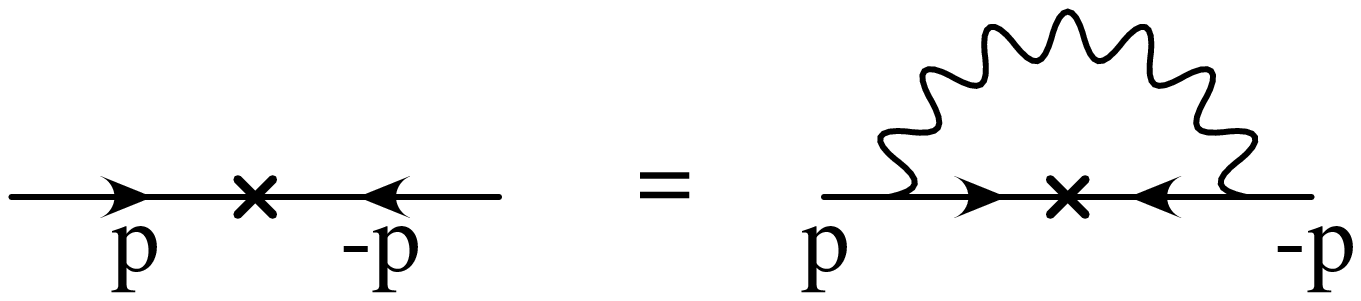,scale=.5}
   \end {center}
   \caption
       {%
          The gap equation for the $\langle qq\rangle$ condensate.
       \label{fig:gap1}
       }%
}
\end {figure}

\begin {figure}
\vbox{
   \begin {center}
      \epsfig{file=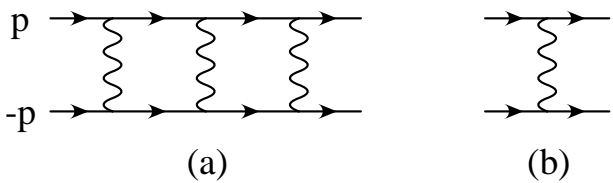,scale=.5}
   \end {center}
   \vspace{-0.2in}
   \caption
       {%
          (a) binding of a Cooper pair by multiple interactions;
          (b) a single such interaction.
       \label{fig:gap2}
       }%
}
\end {figure}

In weak coupling, one can write a self-consistency equation for the
formation of a
condensate $\langle qq \rangle$ of the Cooper pairs.
That equation is known as the gap equation and is
depicted schematically in fig.\ \ref{fig:gap1},
where $\Delta$ represents the condensate (or ``gap'').
If you iterate the gap equation, you can generate rainbows of
gauge interaction lines.  These diagrams represent repeated interactions of
the two fermions, binding them into the Cooper pair, as depicted in
fig.\ \ref{fig:gap2}a.

Let's now look at an individual interaction between the fermions, as in
fig.\ \ref{fig:gap2}b.
This is just Coulomb scattering, and Coulomb scattering is naively infrared
divergent.  This divergence is a manifestation of the long-range nature of
the Coulomb force.
Consider first the case of electric interactions.
Completely analogous to our earlier discussion about high temperature,
electric fields are screened at finite density by the Debye effect
(with the inverse Debye length $m_{\rm D}$ of order $g\mu$).
The Debye effect cuts off the Coulomb divergence for electric interactions.

Now consider magnetic interactions between the two fermions.
Again analogous to high temperature systems, static magnetic fields
are not screened.  In the old days of studying color conductivity, people
simply assumed by analogy with the high-temperature case that there must
be a scale $g^2\mu$ (analogous to $g^2 T$) of non-perturbative magnetic
physics, and that this non-perturbative physics would somehow cut off the
magnetic Coulomb divergence.  I'm told it was understood by a fellow
named Barrois
as early as 1979 that this is incorrect, but he could never get this result
from his Ph.D.
thesis published.  Barrois left the field.
The correct understanding of what cuts off the magnetic
Coulomb divergence
is recent.
In 1998, Son\cite{son}
realized that it is
the screening of {\it dynamical} magnetic fields, due to Lenz's Law,
which I discussed earlier.  Knowing the correct physics, he was able to
then compute the size of the super-conducting gap at zero temperature,
to leading order in the (assumed small) coupling $g(\mu)$:
\begin {equation}
   \Delta = \# g^{-5} \mu \exp\left[-{3\pi^2\over \sqrt{2} \, g(\mu)}\right].
\label{eq:son}
\end {equation}
This formula puts a quantitative face on the assertion that QCD forms a
color superconductor at sufficiently high density.  One would still like to
solve rigorously for what {\it type}\/ (2SC, CFL, {\it etc.}) of superconductor
it is, and that turns out to depend on the numerical coefficient $\#$ in
front of (\ref{eq:son}).  And that turns out to be the outstanding problem
for dense QCD at small coupling: to calculate the numerical prefactor and
so definitively distinguish the fine structure of the phase diagram of
fig.\ \ref{fig:QCDphases2}.%
\footnote{
   A partial but incomplete analysis of this issue has been made by
   Sch\"afer.\cite{schafer}  An analysis of a realted question---the
   corresponding prefactor in determining the critical temperature---
   has been made by Brown, Liu and Ren.\cite{brown}
}

\nonumsection{Acknowledgements}
I would like to thank Krishna Rajagopal, Dam Son, and Thomas Schaefer for
discussions concerning dense QCD that helped me prepare this talk.
This presentation was supported by
the U.S. Department of Energy under Grant No.\ DE-FG02-97ER41027.

\nonumsection{References}
\noindent
I have made no attempt to take on the mammoth task of citing even the
most important works in the field, but I provide references for some
particular statements made in the text.

\end{document}